\documentclass[12pt]{article}
\usepackage[pdftex]{hyperref,graphicx,color}

\newcommand\degree{^{\circ}}
\newcommand{\arcsec}{^{\prime\prime}}
\newcommand{\arcmin}{^{\prime}}
\newcommand{\la}{\,\lower2truept\hbox{${<\atop\hbox{\raise4truept\hbox{$\sim$}}}
$}\,}
\newcommand{\ga}{\,\lower2truept\hbox{${> \atop\hbox{\raise4truept\hbox{$\sim$}}
}$}\,}

\begin{document}

\includegraphics[width=0.2\textwidth]{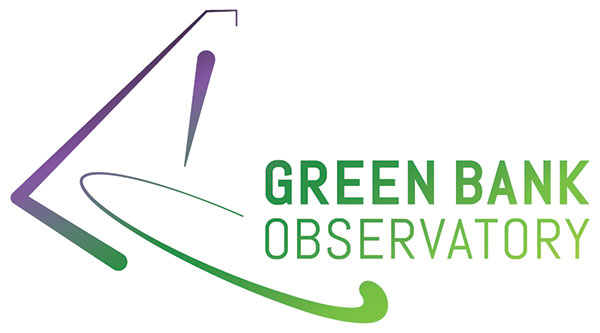}\hspace*{2cm}{\Large{\bf GBT
    Memo \#302}}

\begin{center}
\ \\
{\Large {\bf Calibration of Argus and the 4mm Receiver
    on the GBT}}
\ \\

David T. Frayer, Ronald J. Maddalena, Steven White, \& Galen Watts\\
(Green Bank Observatory)\\
Amanda Kepley (NRAO)\\
Jialu Li \& Andrew I. Harris (University of Maryland)\\

2019 June 05
\end{center}

\begin{abstract}

  The calibration procedures for data collected for Argus and the 4mm
  Receiver instruments on the GBT are presented.  The measured beam
  size, aperture efficiency, and main-beam efficiency are derived for
  the range of observing frequencies (66--116\,GHz) within the 3\,mm
  atmospheric window.  The telescope performs well even at the highest
  frequencies ($>$100\,GHz).  The amount of power in the error pattern
  of the antenna beam is estimated.  Approximately 95\% of the total
  antenna power is contained within $1\degree$ of the pointed
  direction.  The calibration derived using small sources follows
  theoretical expectations based on the Ruze equation.  In general,
  bright point source calibrators that are monitored regularly by ALMA
  can be used to accurately derive the main-beam efficiency of the
  telescope.

\end{abstract}

\section{Introduction}

The 4mm Receiver (Rcvr68\_92) and Argus (RcvrArray75\_115) were
commissioned on the Robert C. Byrd Green Bank Telescope
(GBT)\footnote{The Green Bank Observatory is a facility of the
  National Science Foundation under cooperative agreement by
  Associated Universities, Inc.} in 2012 and 2016 respectively.  The
results from commissioning, measured instrument performance, and
calibration summaries are given on the instrument web
pages\footnote{http://www.gb.nrao.edu/4mm/ and
  http://www.gb.nrao.edu/argus/}.  A short test project was carried
out in 2018 April to verify the relative calibration of the 4mm
Receiver and Argus.  Using these data and the archival calibration
data from science programs, the calibration results for Argus and the
4mm Receiver are presented in this memo.  The calibration of GBT at
lower observing frequencies, which is based on noise-diodes, is
discussed in the GBTIDL spectral-line processing documentation Braatz
(2009).

The table of observations used in this memo is provided in Appendix~A,
and the telescope efficiencies are defined in their integral forms in
Appendix~B.

\section{Temperature Scales and Efficiency Definitions}

The antenna temperature ($T_{A}$) is
related to antenna temperature corrected for atmosphere
($T_{A}^{\prime}$) by
\begin{equation}
T_{A}^{\prime} = T_{A}\,\exp(\tau_{o}\,A),
\end{equation}
where $\tau_{o}$ is the zenith opacity and $A$ is the airmass which
can be approximated by $A\simeq 1.0/\sin({\rm El})$, except at very
low elevations (El).

The forward beam brightness temperature $T_{A}^{*}$ is the effective
beam brightness temperature seeing the entire the sky corrected for
atmosphere.
\begin{equation}
T_{A}^{*} = \frac{T_{A}^{\prime}}{\eta_{l}},
\end{equation}
where $\eta_{l}$ represents the faction of power in the forward $2\pi$
steradian direction on the sky and ($1-\eta_{l}$) is the rear
``spillover'' and scattering fraction that sees the ground.  Following
the convention of Kutner \& Ulich (1981), the value $\eta_{l}$ is a
combination of the radiation efficiency $\eta_{r}$ that corrects for
ohmic losses and the ``rearward'' scattering and spillover efficiency
($\eta_{rss}$), where $\eta_{l} =\eta_{r}\,\eta_{rss}$.  Since
$\eta_{r}\simeq 1.0$ for radio telescopes, $\eta_{l} \simeq
\eta_{rss}$.  Using the definitions adopted for the IRAM 30m (Kramer
1997), $\eta_{l}$ is the forward efficiency (F$_{\rm eff}$).  For the
unblocked aperture, the feed-arm design, and Gregorian geometry of the
GBT, the forward efficiency is expected to be approximately $\eta_{l}
\simeq 0.99$ (Srikanth 1989a,b).

The main-beam efficiency $(\eta_{mb})$ is used
to relate $T_{A}^{\prime}$ to the measured main-beam antenna
temperature ($T_{mb}$) by
\begin{equation}
T_{mb}= \frac{T_{A}^{\prime}}{\eta_{mb}}.
\end{equation}
The main-beam efficiency represents the fraction of power contained
within the main-beam lobe of the antenna pattern.

Kutner \& Ulich (1981) introduced $T_{R}^{*}$ which corrects for both
forward and rearward losses.
\begin{equation}
T_{R}^{*}= \frac{T_{A}^{\prime}}{\eta_{l}\,\eta_{fss}},
\end{equation}
where $\eta_{fss}$ is called the ``forward'' scattering and spillover
efficiency.  The value $\eta_{fss}$ is the faction of power in the
forward $2\pi$ direction that is contained within the diffraction
pattern of the telescope including the error beam pattern (Appendix
B).  The corrected main-beam efficiency ($\eta_{M}^{*}$) relates the
$T_{R}^{*}$ temperature scale to the main-beam $T_{mb}$ temperature
scale (Mangum 1999);
\begin{equation}
\eta_{M}^{*} = \frac{T_{R}^{*}}{T_{mb}} = \frac{\eta_{mb}}{\eta_{l}\,\eta_{fss}}.
\end{equation}

Single-dish astronomers typically report observational results using
temperature scales of either $T_{A}^{\prime}$, $T_{mb}$, $T_{a}^{*}$,
or $T_{R}^{*}$, depending on the calibration practices at individual
telescopes and their scientific needs.  These temperature scales are
not the same, and are related by the following expressions:
\begin{equation}
T_{A}^{\prime}\,=\,\eta_{mb}\,T_{mb}\,=\,\eta_{l}\,T_{A}^{*}\,=\,\eta_{l}\,\eta_{fss}\,T_{R}^{*}.
\end{equation}

\subsection{Small Sources}
For sources much smaller than the telescope beam, the aperture efficiency
($\eta_{a}$) of the telescope relates the flux density
($S_{\nu}$) of the source to $T_{A}^{\prime}$.  For the 100m GBT, the
aperture efficiency is
\begin{equation}
\eta_{a}= 0.352\frac{T_{A}^{\prime}}{S_{\nu}}.
\end{equation}
Assuming the Ruze equation, the aperture efficiency is related to the
surface errors of the telescope by:
\begin{equation}
\eta_{a} =\eta_{0} \exp[-(4\pi\epsilon/\lambda)^2)],
\end{equation}
where the coefficient $\eta_{0} = 0.71$ is the GBT aperture efficiency
at long wavelengths, $\epsilon$ is the rms uncertainty of the surface,
and $\lambda$ is the observing wavelength.  Currently for optimal GBT
surface corrections and under excellent observing conditions,
$\epsilon \simeq 230$\,$\mu$m (Frayer et al. 2018).

Assuming a Gaussian beam and for point sources, the
main-beam efficiency is given by
\begin{equation}
\eta_{mb} = 0.8899\eta_{a}\left(\theta_{FWHM}\frac{D}{\lambda}\right)^2,
\end{equation}
where $\theta_{FWHM}$ is the full-width half-maximum beam size in
radians and D is the 100m diameter of the GBT (Maddalena 2010).

\subsection{Extended Sources}

For sources much larger than the telescope beam, the source brightness
temperature ($T_{S}$) is well approximated by the main-beam
temperature, i.e., $T_{mb} \simeq T_{S}$.  However, measurements of
extended sources includes power on spatial scales that resides outside
of the main-beam lobe.  The measured antenna temperature depends on
the coupling of the source structure with the full the antenna beam
pattern, including the error pattern.  Defining the efficiency with
which the antenna couples to the source as $\eta_{c}$ (Appendix B),
$T_{A}^{\prime} = \eta_{c}\eta_{fss}\eta_{l}\,T_{S}$.  This definition
is consistent with Kutner \& Ulich (1981) where $T_{R}^{*} =
\eta_{c}\,T_{S}$.  The efficiency terms can be combined to define the
effective beam efficiency for a source as
\begin{equation}
\eta_{\rm Source} = \eta_{c}\eta_{fss}\eta_{l},
\end{equation}
such that for an extended source
\begin{equation}
T_{A}^{\prime} = \eta_{\rm Source}\,T_{S}.
\end{equation}

\section{Calibration Equations} 

The equations associated with the calibration of the 4mm Receiver were
previously presented in the 67-93.6 GHz Spectral Line Survey paper for
Orion-KL (Frayer et al. 2015).  The equations for Argus are based on
Kutner \& Ulich (1981) and are summarized on the Argus web page.  The
4mm Receiver has both a cold and ambient load (``two load'' system)
while Argus only has an ambient load (``one load'' system) so the
calibration approaches of the two instruments are slightly different.
Both approaches are summarized here.

\subsection{The 4mm Receiver}

The 4mm Receiver has two feeds that are separated by a cold load within the
cryostat.  A calibration wheel rotates above the cryostat which permits
the feeds to see the sky, the ambient load, or mirrors that enable
the feeds to look back down into the cryostat at the cold load.
Using both loads for the 4mm Receiver, the gain ($g$) is given by
\begin{equation}
g = [(T_{amb}-T_{cold})/(C_{amb} - C_{cold})],
\end{equation}
where $T_{amb}$ and $T_{cold}$ are the temperatures of the ambient and
cold loads and $C_{amb}$ and $C_{cold}$ are the observed counts of the
ambient and cold loads.  The temperature of ambient load is measured
by a temperature sensor in real-time, while the effective temperature
of the cold load has been measured in the lab using an external liquid
nitrogen load.  At the default frequency of 77 GHz for the Auto
observing procedures for the 4mm Receiver, the cold load is estimated
to be 54\,K.  Lab measurements suggest the effective cold load
temperature decreases roughly linearly from 60\,K at 67\,GHz to about
45\,K at 92\,GHz.  The cold load temperature in units of Kelvin is
expressed as a function of frequency by
\begin{equation}
T_{cold} = 54.0\,{\rm K} - 0.6{\rm[K/GHz]}(\nu - 77.0\,{\rm GHz}),
\end{equation}
where $\nu$ is the observing frequency in units of GHz.  By using this
relationship, the uncertainty on $T_{cold}$ is estimated to be about
4\,K which corresponds to an uncertainty on the gain and the measured
$T_{A}$ temperature scale of less than 2\%.

For most GBT observations, the (ON$-$OFF)/OFF processing method is
adopted.  Using this approach, the observed antenna temperature is
\begin{equation}
T_{A} = T_{\rm sys} (C_{ON} - C_{OFF})/C_{OFF},
\end{equation}
where  $C_{ON}$ and
$C_{OFF}$ are the observed counts of the ON and OFF scans. 
The system temperature ($T_{\rm sys}$) is given by 
\begin{equation}
  T_{\rm sys} = g\,C_{OFF}, 
\end{equation}
where the gain is given by equation (12).  For the default calibration
procedures, both $T_{\rm sys}$ and $g$ are scalar values representing
the median or average across the central region of the spectral
bandpass.  The observed antenna temperature is corrected for
atmosphere ($T_{A}^{\prime}$) using equation (1).  The opacity is
derived from the Observatory's weather database derived for Green
Bank.  Except for periods of rapidly changing weather conditions, the
predicted opacities are accurate to within $\Delta \tau_o \simeq 0.01$
based on historical measurements.

\subsection{Argus}

The Argus instrument is a 16 element focal-plane-array operating from
74\,GHz to 116\,GHz.  It has a chopper-vane absorber that can be
placed over the array for calibration.  For the chopper-vane technique
with only one load, the natural temperature scale measured by the
antenna is $T_{A}^{*}$ (Kutner \& Ulich 1981; Kramer 1997).  This
temperature scale has the advantage of being corrected for atmospheric
attenuation while its derivation is nearly independent of opacity.

Analogous to equation (12), the ``gain'' for a one load system is
derived from the difference of signals measured between the ambient
load and blank sky.

\begin{equation}
g = [(T_{amb}-T_{OFF})/(C_{amb} - C_{OFF})],
\end{equation}
where $T_{amb}$ and $T_{OFF}$ are the temperatures of the ambient load
and the blank sky (OFF) and $C_{amb}$ and $C_{OFF}$ are the
corresponding observed counts of the ambient load and blank sky.
The $T_{A}^{*}$ temperature is
\begin{equation}
 T_{A}^{*} = T_{cal} (C_{ON} - C_{OFF})/ (C_{amb} - C_{OFF}),
\end{equation}
where 
\begin{equation}
 T_{cal} = (T_{amb}-T_{OFF})\exp(\tau_{o}\,A)/\eta_{l}.
\end{equation}
The difference $T_{amb}-T_{OFF}$ removes the
receiver temperature and other telescope systematics, and is given by 
\begin{equation}
(T_{amb} - T_{OFF}) = T_{amb} - (\eta_{l}\,T_{sky} + (1-\eta_{l})\,T_{spill})
\end{equation} 
where $\eta_{l}\,T_{sky}$ is the
contribution from the sky and $(1-\eta_{l})\,T_{spill}$ is
the contribution from ground spillover.
The temperature of the sky is
\begin{equation}
T_{sky}= T_{ATM}(1-\exp(-\tau_{o}\,A)) + T_{bg}\exp(-\tau_{o}\,A),
\end{equation}
where $T_{ATM}$ is the effective atmospheric temperature and $T_{bg}$
is the cosmic microwave background temperature of 2.73\,K.  With a
little algebra, the equations can be reduced to
\begin{equation}
T_{cal} = (T_{ATM} - T_{bg}) + (T_{spill} - T_{ATM})\exp(\tau_{o}\,A)
+ (T_{amb}-T_{spill})\exp(\tau_{o}\,A)/\eta_{l}.
\end{equation}
Equation (21) is the same as equation (8) of Kramer (1997) and
equation (12) of Kutner \& Ulich (1981).  If $T_{amb} \simeq
T_{spill}$ or if $\eta_{l} \simeq 1$ (which are both good
approximations for the GBT), then
\begin{equation}
T_{cal} \simeq (T_{ATM}-T_{bg})+(T_{amb}-T_{ATM})\exp(\tau_{o}\,A).
\end{equation}

Based on the 2017/2018 observing season with Argus on the GBT,
the average $T_{cal}/T_{amb}$ ratio using equation (22) is
$0.997\pm0.018$.  Hence, one could approximate the calibration by
simply assuming
 \begin{equation}
T_{cal} \approx T_{amb}.
\end{equation}
For example, if one adopts typical values ($T_{bg}=2.7$\,K,
$\exp(\tau_{o}\,A)=1.2$, $\eta_{l} \simeq 0.99$, $T_{ATM}=260$\,K,
$T_{spill}=270$\,K, $T_{amb}=273$\,K), one finds that equations (21,
22, \& 23) all yield the same value of $T_{cal}\simeq 273$\,K.  This
approximation works best for average observing conditions.  In the
coldest conditions $T_{amb} < 260$\,K, equation (23) tends to
overestimate $T_{cal}$ by about 0.7\%, and the $T_{cal} \approx
T_{amb}$ approximation begins to break down with large atmospheric
opacities ($\exp[\tau_{o}\,A]>1.7$) where $T_{cal}$ is underestimated
by 5-10\%.

For Argus observations, the chopper-vane calibration is done at least
every 30-45 minutes which is the recommended time between pointing and
focus observations.  The measured signal from the vane has been shown
to be stable over a time scale of more than one hour in good
conditions.  However, the effective system temperature can change
rapidly with varying sky conditions as a function of time.
To correct for the varying effective system temperature and analogous
to equation (14),

\begin{equation}
T_{A}^{*} = T_{\rm sys}^{*} (C_{ON} - C_{OFF})/C_{OFF},
\end{equation}
where  $C_{ON}$ and $C_{OFF}$ are the observed counts of the ON and
OFF scans, and the effective system temperature ($T_{\rm sys}^{*}$) as
a function of time ($t$) is given by 
\begin{equation}
  T_{\rm sys}^{*}(t) = T_{cal}/(C_{amb}/C_{OFF}(t) -1).
\end{equation}
The effective system temperature is related to $T_{sys}$ by
\begin{equation}
  T_{\rm sys}^{*} = T_{sys}\exp(\tau_{o}\,A)/\eta_{l}.
\end{equation}
Equations (24) and (25) are used to calibration the raw counts into
$T_{A}^{*}$ for the Argus VEGAS (spectrometer) data, with $T_{cal}$
given by equation (22).  For the default calibration procedures,
$T_{\rm sys}^{*}$ and $T_{cal}$ are scalar values representing the
median or average across the central region of the spectral
bandpass. For the pointing and focus data taken with the Digital
Continuum Receiver (DCR), equation (23) is adopted for $T_{cal}$ when
reporting system temperatures within the Astrid observing window.

\subsection{Equation Simplifications}

For simplicity, the equations in \S3 adopt the Rayleigh-Jeans
approximation, i.e., $J(T) = T$, where ($h\,\nu << k\,T$).  The
physical temperatures should be replaced by the Planck function
radiation temperature ($J_{\nu}(T_{b}$)) for accurate
calculations.   The Planck function radiation temperature is
\begin{equation}
J_{\nu}(T_{b}) =
\frac{h\,\nu}{k}\left(\exp\left[\frac{h\,\nu}{k\,T_{b}}\right]
  -1\right)^{-1},
\end{equation}
where $T_{b}$ is the brightness temperature.  

Another simplification to the standard calibration equations arises
since both the 4mm Receiver and Argus are single side-band systems.
For single side-band receivers, the signal from the image side-band is
negligible such that such that $G_{i}/G_{s} \simeq 0$, for comparison with
the equations of Kutner \& Ulich (1981) and Kramer (1997).

\section{Results}
 
The calibration of Argus and the 4mm Receiver is tied to the ALMA
calibration system using the online ALMA Calibrator Source Catalog
(https://almascience.eso.org/sc/).  This web-tool is used to derive
the flux densities of bright 3mm quasars for absolute flux calibration
of Argus and the 4mm Receiver.  The cataloged ALMA Band-3 measurements
at 91.5 GHz and 103.5 GHz are extrapolated to the observed GBT
frequency using the inferred Band-3 spectral-index and interpolated in
time between ALMA measurements surrounding the GBT observations.  ALMA
regularly observes all bright 3mm quasars that are visible at the ALMA
site (23$\degree$ South), including even 3C84 which is at a
declination of $+41.5\degree$.

In addition to using bright 3mm quasars, observations of Jupiter,
Mars, and the Moon were used to measure the beam efficiency associated
with different source sizes.  The forward efficiency $\eta_{l}$ is
estimated to be 0.985, and this value is adopted for this memo.
Accurate absolute measurements of $\eta_{l}$ are challenging to
measure at 3mm on the GBT due to weather dependencies and since its
value is near 1.  Based on tipping observations and comparisons
between $T_{A}^{*}$ and $T_{A}^{\prime}$, $\eta_{l} \simeq 0.985 \pm
0.015$.  This is consistent with the theoretical expectations for the
GBT of $\eta_{l} \simeq 0.99$ (Srikanth 1989a; Srikanth 1989b).
Possible variations of $\eta_{l}$ as a function of elevation are not
included in this memo.

\subsection{Comparison Between the 4mm Receiver and Argus}

Observations were carried out on 2018 April 06 to compare the relative
calibration of the 4mm Receiver and Argus (GBT project and session:
TGBT17B\_502\_01).  No significant difference in the calibration of
these instruments were found using their respective calibration
approaches outlined in \S3.  Observations of the bright quasar 3C279
were used to measure the aperture efficiencies, while observations of
Jupiter and the Moon were used to derive the beam efficiencies for
these extended sources.

The aperture efficiencies were derived using equation (7).  The beam
efficiencies associated with extended sources were derived using
\begin{equation}
  \eta_{\rm Source} = \frac{T_{A}^{\prime}}{J_{\nu}(T_{b})} =
  \frac{\eta_{l}T_{A}^{*}}{J_{\nu}(T_{b})}.
\end{equation}
Since Jupiter and the Moon are much larger than the beam size,
equation (11) is applicable.  The brightness temperature ($T_{b}$) as
a function of frequency for the planets were taken from the CASA
software package (Butler 2012).  The radiation temperature of the Moon
was computed based on the phase of the Moon at the time of
observations using the formulae presented in Mangum (1993).

Figure~1 shows the comparison of the measurements.  The values derived
for 3C279, Jupiter, and the Moon were all consistent within
measurement errors between the 4mm Receiver and Argus.  The measured
beam efficiency for the Moon is larger than that observed for Jupiter
which implies significant power within the beam pattern on spatial
scales between $\sim 40\arcsec$ (size of Jupiter) and $32\arcmin$
(size of the Moon).  The measured efficiencies were not optimal as the
weather was somewhat marginal for calibration observations (moderate
opacities that varied significantly throughout the observing session).
The dashed-line shows the expected aperture efficiency based on an
effective surface uncertainty of 235$\mu$m, which is the best-fit
average value for quality observations during the 2017 and 2018
observing seasons (\S4.2).

\subsection{Small Source Calibration}

Approximately 200 measurements of point sources were collected using
observations with the 4mm Receiver and Argus from 2017 February
through 2019 March to quantify the performance of the GBT within the
3\,mm atmospheric window (Appendix A).  Only observations of the
brightest point-source calibrators were used ($S{\nu} > 3$\,Jy), and
sessions with questionable pointing, focus, or surface corrections
were discarded.  The sample used for this analysis is not complete,
but is representative of the wide range of conditions over the full
high-frequency observing season (October through April) in Green Bank.

\subsubsection{Beam Size}

Figure~2 shows the measured beam sizes as a function of frequency.
The error bars represented the measurement uncertainty of individual
measurements divided by the square-root of the number of observations
within each frequency bin.  Most observations, to date, have been done
within the 86--90 GHz frequency band corresponding to HCN(1-0) and
HCO$^{+}$(1-0) and that of $^{13}$CO near 110 GHz.  The solid line in
Figure~2 corresponds to
\begin{equation}
\kappa \equiv \theta_{\rm FWHM} \left(\frac{D}{\lambda}\right) = 1.20,
\end{equation}
whose normalization is derived from the best fit parameter of
$\kappa=1.20\pm0.02$ over all frequencies.  Under typical conditions,
the beam performance degrades slightly at the highest frequencies.
This is shown by the data point lying above the line in Figure~2 at
frequencies near 110 GHz.

The effective $\kappa$ for individual observing sessions range from
about 1.15 to 1.3, depending on the winds and the derived telescope
corrections.  Under very good conditions, one can achieve $\kappa <
1.18$ even at high-frequency.  For example, a value of $\kappa = 1.16$
was derived at 109.4\,GHz in calm conditions and with very low
opacity.  This result is discussed in GBT Memo\#296 (Frayer 2017) and
is shown by the open circle in Figure~2.

\subsubsection{Aperture and Main-Beam Efficiencies}

For the sample of bright point-sources, the aperture efficiencies were
derived using equation~(7).  The results as a function of frequency
are consistent with the Ruze equation (equation 8) with an effective
surface error of 235$\mu$m.  This value represents all losses (e.g.,
pointing errors, focus errors, tracking errors due to winds and
telescope servo system, as well as surface errors from the dish and
subreflector).  The main-beam efficiency was derived from the
measurements of the peak scans of the pointing sources using
equation~(9).  The uncertainties on the individual observations are
large due to measurement errors.  The typical measurement
uncertainties on $\eta_{a}$ and $\theta_{FWHM}$ result in a 23\%
measurement error on $\eta_{mb}$.  The uncertainties plotted in
Figure~3 represent the measurement uncertainty of individual
observations divided by the square-root of the number of observations
within each frequency bin.  The data points near 88 GHz and 110 GHz
are the most accurate due to the large number of observations made at
these frequencies.

From the derived efficiencies and errors within each frequency bin, the weighted
average ratio derived is
\begin{equation}
\frac{\eta_{mb}}{\eta_{a}} = 1.274\pm0.035.
\end{equation} 
This ratio is consistent with the average beam size parameter of
$\kappa =1.20$ derived in \S4.2.1, assuming a Gaussian beam.  The
$\eta_{mb}$ data point measured near 110\,GHz is higher than would be
implied by equation 30, which results from the slightly larger average
value of $\kappa$ at this frequency as shown in Figure~2.

\subsection{Corrected Main Beam Efficiency}

The corrected main-beam efficiency $\eta_{M}^{*}$ can be inferred from
observations, but since it depends significantly on the error beam
pattern, theoretical considerations are typically used to estimate its
value.  Using the relationships associated with the Ruze equation (see
discussion in Baars 1973 and the application for the NRAO 12m in
Mangum 1999),
\begin{equation}
\eta_{M}^{*} = \left(1 + \frac{A_{E}\theta_{E}^{2}}{A_{M}\theta_{M}^{2}}\right)^{-1},
\end{equation}
where $A$ and $\theta$ are the amplitudes and half-power width,
respectively, corresponding to the error ($E$) and main ($M$) beam
patterns.  

Assuming a Gaussian beam and a Gaussian error pattern, the
value $(A_{E}\theta_{E}^{2})/(A_{M}\theta_{M}^{2})$ represents the
ratio of power in the error pattern compared to the main beam.
The FWHM spatial scale of the error pattern is
\begin{equation}
\theta_{E} = \frac{2}{\pi}\sqrt{\ln(2)}\left(\frac{\lambda}{c_{\sigma}}\right),
\end{equation}
where $c_{\sigma}$ is the surface deviation correlation radius.  The
ratio of the amplitude of the error beam relative to the amplitude of
the main beam is given by
\begin{equation}
\frac{A_{E}}{A_{M}} =
\frac{1}{\eta_{0}}\left[\frac{2\,c_{\sigma}}{D}\right]^2\,(\exp(\delta^2)-1),
\end{equation}
where $\delta = 4\pi\,\epsilon/\lambda$. 

Based on the above equations, $\eta_{M}^{*}$ and $\eta_{fss}$ can be
computed as a function of frequency for the GBT (Figure~4).  These
derived efficiencies are not sensitive to the value $c_\sigma$ since
it cancels out in equation (31).  The above equations reduce to
\begin{equation}
\eta_{M}^{*} = \left(1 +
\frac{16\ln(2)}{\pi^{2}\kappa^{2}}\left[\frac{1}{\eta_{a}} - \frac{1}{\eta_{0}}\right]\right)^{-1}.
\end{equation}
This is a useful general expression that only depends on the beam
parameter ($\kappa$) and the aperture efficiency, which is given by
the surface rms $\epsilon$.  Equations~(32) and (33) are derived
assuming $\delta < 1$, i.e., at wavelengths for which the telescope
performs well.  In the case of $\delta > 1$, the additional terms
required in the analysis cancel out when computing $\eta_{M}^{*}$,
allowing equation~(34) to be useful for all values of $\delta$.

The forward spillover and scatter efficiency can also be expressed
theoretically based only on the telescope $\kappa$ and $\epsilon$
parameters;
\begin{equation}
\eta_{fss} = \frac{0.8899\kappa^{2}\eta_{a}}{\eta_{l}\eta_{M}^{*}}.
\end{equation}
At 86\,GHz the derived efficiencies are $\eta_{M}^{*} = 0.465\pm0.035$
and $\eta_{fss}=0.965\pm0.020$, adopting $\kappa=1.2$ and
$\epsilon=235\mu$m.

\subsection{Antenna Pattern}

By assuming a Gaussian approximation for the main-beam and the error
pattern and using the equations in the previous sub-section, the
physical scale ($c_{\sigma}$, correlation radius) associated with the
error pattern can be estimated.  Baars (1973) discusses the method for
deriving this parameter based on the measured beam efficiencies of
small sources and a large extended source.  The correlation radius is
given by
\begin{equation}
c_{\sigma} =
\frac{2\lambda}{\pi\theta_{S}}\left[-\ln\left(\frac{\eta_{B0} 
-\eta_{B}^{\prime}}{\eta_{B0}-\eta_{mb}}\right)\right]^{0.5},
\end{equation}
where the main-beam efficiency $\eta_{mb}$ is given by equation (9)
and $\eta_{B0}$ is the idealized main-beam efficiency at long
wavelengths ($\eta_{B0}\simeq 0.905\,[0.71\times1.274]$ for the
GBT). The value $\eta_{B}^{\prime} $ is the beam efficiency
measured for an extended source with angular size $\theta_{S}$, where
$\theta_{mb} < \theta_{S} < \theta_{E}$.

Using measurements of Jupiter ($\theta_{S}=43\arcsec$) at 86--89\,GHz,
$\eta_B^{\prime} = \eta_{\rm Jupiter} = 0.53$.  At the default Argus
calibration frequency of 86\,GHz, $\eta_{a} = 0.347$ and $\eta_{mb} =
0.442$ (assuming typical 235$\mu$m surface errors).  Based on these
values and using equation (36), a value of $c_\sigma \simeq 500$\,cm
is derived, which corresponds to an angular scale of $76\arcsec$.  In
comparison, the GBT panels are $\sim 250 \times 200$\,cm in size
(Schwab \& Hunter 2010), so the derived correlation radius corresponds
to a size of about 2 panels.  The Gravity Zernike model and the
AutoOOF Thermal Zernike coefficients correct for significantly larger
scale distortions of order 20m across the dish.  This suggest that the
correlation radius should be fairly insensitive to the Zernike
coefficients.

By using observations of the Moon, the amount of power on larger
spatial scales outside of the error pattern measured with Jupiter can
be determined.  Measurements of the Moon using Argus yield $\eta_{\rm
  Moon} = 0.81\pm0.038$ at 88.9\,GHz and $0.82\pm0.044$ at 110.0 GHz.
There is no evidence for any significant frequency dependency with
$\eta_{\rm Moon}$ within the 3mm band.  Based on the weighted average
of these observations, a value of $\eta_{\rm Moon} = 0.814\pm0.029$ is
adopted.  This value is significantly lower than $\eta_{fss}$ and what
would be expected by integrating over the main-beam pattern and the
error pattern implied by the observations of Jupiter (dashed-dotted
line in Figure~5).  This suggests an additional error component to the
beam pattern.  The Jupiter and Moon data are fitted with a double
Gaussian error pattern with the results shown in Figure~5 and Table~1.
The data were fitted using three free parameters, which are the size
scales associated with each of the two components of the error pattern
($c_{E1}=700\pm200$\,cm and $c_{E2}=20\pm10$\,cm) and the relative
amount of power contained within the two components
($A_{E1}\theta_{E1}^{2}/(A_{E2}\theta_{E2}^{2})=1.0\pm0.5$).

\subsection{Extended Source Calibration}

The planets can be used to derive beam efficiencies of the telescope.
The main-beam efficiency measured for a planet of known brightness
temperature $T_{b}$ and diameter $\theta_{S}$ is given by
\begin{equation}
\eta_{mb} = \frac{T_{a}^{\prime}}{J_{\nu}(T_{b})[1-\exp(-x^{2})]},
\end{equation}
where $x = \sqrt{\ln\,2}(\theta_{S}/\theta_{mb})$ (e.g., Kramer 1997).
This equation is useful for the small angular-size planets, such as
Mars and Uranus which are the two primary calibrator sources used for
mm astronomy.  The value of $x$ assumes a Gaussian telescope beam,
which is a good approximation for planet disks smaller than
$\theta_{mb}$.  For disks $\theta_{S}/\theta_{mb} \ga 1$, then using a
Bessel function for telescope beam is more appropriate than a
Gaussian.  For very large sources $\theta_{S}/\theta_{mb} > 3$, such
as Jupiter or the Moon, then approximation given by equation (11) is
applicable.

Figure~6 compares the measured main-beam efficiencies ($\eta_{mb}$)
for quasars to the beam efficiency measured for Jupiter ($\eta_{\rm
  Jupiter}$).  The values of $\eta_{\rm Jupiter}$ are larger than
$\eta_{mb}$ given that measurements of Jupiter include significant
power from the antenna error pattern (\S4.4).  For small sources,
variations in $\eta_{a}$ directly correlate with variations in
$\eta_{mb}$ (equation 9) as expected for different observing sessions,
depending on the conditions and the quality of the telescope
corrections.  Since Jupiter is much larger than the beam size, one
could expect the measured $\eta_{\rm Jupiter}$ values to be fairly
independent of the detailed telescope corrections.  However, we found
significant variations of $\eta_{\rm Jupiter}$ that correlated roughly
linearly with $\eta_{a}$.  The solid squares for Jupiter in Figure~6
were derived in sessions with good aperture efficiencies consistent
with the adopted $\epsilon =235\mu$m derived here and used in
\S4.2--4.4.  For session TGBT17B\_502\_01, both $\eta_{\rm Jupiter}$
and $\eta_{a}$ were 10\% lower, and for session AGBT18B\_370\_02 both
$\eta_{\rm Jupiter}$ and $\eta_{a}$ were about 20\% lower than the
nominal values.  After scaling by the relative aperture efficiencies
between sessions, the $\eta_{\rm Jupiter}$ values are consistent
between sessions (open circles and solid squares in Figure 6).  This
plot potentially embeds information on the telescope error beam as a
function of frequency (following the analysis in \S4.4).  However,
additional well-calibrated scanning observations of Jupiter and the
Moon are needed to more accurately constrain the antenna pattern of
the GBT at high frequency.

\section{Discussion}

Table~2 shows the derived GBT performance values at 86\,GHz.  This
memo presents relationships to compute $\eta_{a}$, $\eta_{mb}$,
$\eta_{M}^{*}$, and $\eta_{fss}$ as a function of frequency based only
on the effective surface error $\epsilon$ and the beam size $\kappa$
parameter.  These relationships assume the dominate effects on
performance of the telescope are related to surface deformations using
the Ruze equations.  In cases of low inferred aperture efficiencies
that are due to pointing and/or focus errors, for example, these
relationships that depend on $\eta_{a}$ may not be applicable.

There is a fairly wide-range of derived aperture efficiencies for the
GBT at 3\,mm depending on the weather conditions and the quality of
the telescope corrections.  Sessions with unreasonably low source
amplitudes were not used in this analysis.  For the 143 observations
from 86--90\,GHz presented here, $\eta_{a}=0.33\pm0.05$, but it is not
unusual to measure values as low as $\eta_{a}\simeq 0.2$ from archival
data.  The analysis of the archival data has highlighted the
challenges for users, at times, to obtain quality data.  The default
pointing/focus/OOF software solutions are not always appropriate, and
software updates could improve the ability to derive reasonable
solutions in less than ideal weather conditions.  Winds can be
particularly troublesome during the OOF observations.  In these cases,
the derived OOF corrections could significantly decrease the
performance of the telescope.  It is recommended to use the Ka+CCB
system for deriving the OOF corrections, when it is available on the
telescope, since winds have a smaller effect at Ka-band frequencies
(30 GHz) given the larger beam size, and the Ka+CCB system provides
much higher signal-to-noise measurements than can be achieved at
W-band frequencies.

For measuring efficiencies, more reliable results are obtained by
slewing the over the source instead of using the NOD or ON$-$OFF
observing routines.  By slewing over the source, the background sky
signal as a function of time is more accurately determined.  For small
sources, the aperture efficiency is given by the peak signal of the
scan, while the main-beam efficiency is derived by integrating over
the source profile contained within the main-beam.

We recommend absolute calibration to be carried out on bright quasars,
taking advantage of the on-line ALMA calibrator catalog.  Planets can
also be used, but the size of the planet should be less than the beam
size to avoid complications in estimating the effects of source
coupling with the antenna beam pattern.  In terms of brightness and
size, only Mars is suitable for main-beam efficiency measurements
(when Mars is not near the Earth).  Measurements of Jupiter and the
Moon can be used to constrain the error patterns of the telescope beam.

\section{References}

\begin{itemize}

\item Baars, J. W. M., 1973, The Measurement of Large Antennas with
  Cosmic Radio Sources, IEEE Tran. on Antennas and Propagation, Vol
  AP-21, No. 4, 461

\item Braatz, J. 2009, Calibration of GBT Spectral Line Data in GBTIDL v2.1

\item Butler, B. 2012, Flux Density Models for Solar System Bodies in
  CASA, ALMA Memo \#594

\item Frayer, D. T. 2017, The GBT Beam Shape at 109 GHz, GBT Memo
  \#296

\item Frayer, D. T., Ghigo, F., \& Maddalena, R. J. 2018, The GBT Gain
  Curve at High Frequency, GBT Memo \#301



\item Frayer, D. T., et al. 2015, The GBT 67--93.6 GHz Spectral Line
  Survey of Orion-KL, AJ, 149, 162
 
\item Kramer, C. 1997, Calibration of spectral line data at the IRAM
  30m radio telescope

\item Kutner, M. L. \& Ulich, B. L. 1981, Recommendations for
  calibration of millimeter-wavelength spectral line data, ApJ, 250, 341

\item Maddalena, R. J. 2010, Theoretical Ratio of Beam Efficiency and
  Aperture Efficiency, GBT Memo \# 276

\item Mangum, J. G. 1993, Main-beam efficiency measurements of the
  Caltech Submillimeter Observatory, PASP, 105, 117

\item Mangum, J. G. 1999, Equipment and Calibration Status for the
  NRAO 12 Meter Telescope

\item Schwab, F.R. \& Hunter, T.R. 2010, Distortions of the GBT Beam
  Pattern Due to Systematic Deformations of the Surface Panels, GBT
  Memo \#271

\item Srikanth, S. 1989a, Spillover Noise Temperature Calculations for the
Green Bank Clear Aperture Antenna, GBT Memo \#16

\item Srikanth, S. 1989b, More on the Spillover Noise Temperature Comparisons
for the Green Bank Telescope GBT Memo \#19

\item Wilson, T.~L., Rohlfs, K., \& H\"{u}ttemeister, S. 2010, Tools
  of Radio Astronomy, Fifth Ed. (Springer-Verlag Berlin Heidelberg)

\end{itemize}

\section{Tables}

\begin{table}[h!]
\centering
  \caption{86GHz Beam Pattern Measurements}
  \begin{tabular}{lccc}
    \hline \hline
& Relative Amplitude&$\theta$(FWHM) &Fractional Power$^{a}$\\
\hline
MB& 1.0   & 8.64 arcsec& 0.464\\
E1& $0.014\pm0.004$ & $55\pm16$ arcsec & $0.268\pm0.134$\\
E2& $(1.2\pm0.6)\times 10^{-5}$ & $1900\pm950$ arcsec & $0.268\pm0.134$\\
\hline
 \end{tabular}
\begin{flushleft}
  $^{a}$Fractional power does not include the estimated 5\% of total
  power that resides outside of the main-beam (MB) and error
  components (Fig. 5), which is given by $(1 - \eta_{fss}\,\eta_{l})$.
\end{flushleft}
\end{table}

\begin{table}[h!]
\centering
  \caption{86GHz GBT Efficiency and Calibration Parameters}
  \begin{tabular}{lcc}
    \hline \hline
Dish Diameter\dotfill & D & 100 m\\
RMS Surface Accuracy\dotfill & $\epsilon$ & $235\pm15\,\mu$m\\
Beam Size Parameter\dotfill & $\kappa$ & $1.20\pm0.02$\\
Aperture Efficiency\dotfill & $\eta_{a}$ & $0.347\pm0.032$\\
Main-Beam Efficiency\dotfill & $\eta_{mb}$ &$0.442\pm0.043$\\
Corrected Main-Beam Efficiency\dotfill & $\eta_{M}^{*}$&$0.465\pm0.035$\\
Jupiter Beam Efficiency($43^{\prime\prime}$diameter)&$\eta_{\rm Jupiter}$ & $0.53\pm0.05$ \\
Moon Beam Efficiency ($32^{\prime}$ diameter) & $\eta_{\rm Moon}$ &
$0.814\pm0.029$\\
Rear Spillover Efficiency$^{a}$\dotfill & $\eta_{l}$ & $0.985\pm0.015$\\
Forward Spillover Efficiency$^{b}$\dotfill & $\eta_{fss}$ & $0.965\pm0.020$\\

\hline
 \end{tabular}
\begin{flushleft}
  $^{a}$Power in the forward $2\pi$ direction.  $^{b}$Factional power
  in the forward direction inside the $\sim 1^{\circ}$ diameter error
  pattern.
\end{flushleft}
\end{table}

\clearpage
\section{Appendix A}

The listing of GBT observing sessions used in the
data analyses is provided.

\begin{table}[h!]
\centering
  \caption{GBT Observing Sessions}
  \begin{tabular}{lcc}
    \hline \hline
GBT Project    & Session& Date\\
\hline
AGBT17A\_304   & 02     & 2017.02.15\\
AGBT17A\_304   & 03     & 2017.02.27\\
AGBT17A\_304   & 05     & 2017.03.13\\
AGBT17A\_304   & 06     & 2017.03.23\\
AGBT17A\_304   & 07     & 2017.04.09\\
AGBT17A\_304   & 08     & 2017.04.10\\
AGBT17A\_304   & 09     & 2017.04.11\\
AGBT17B\_151   & 02     & 2017.10.16\\
AGBT17B\_151   & 03     & 2017.10.19\\
AGBT17B\_151   & 04     & 2017.11.12\\
AGBT17B\_151   & 05     & 2017.11.14\\
AGBT17B\_151   & 07     & 2017.11.18\\
AGBT17B\_151   & 12     & 2017.12.15\\
AGBT17B\_151   & 13     & 2017.12.21\\
AGBT17B\_151   & 15     & 2018.01.16\\
AGBT17B\_151   & 18     & 2018.01.26\\
AGBT17B\_151   & 19     & 2018.01.27\\
AGBT17B\_151   & 20     & 2018.01.31\\
AGBT17B\_151   & 23     & 2018.02.03\\
AGBT17B\_151   & 24     & 2018.02.26\\
AGBT17B\_151   & 25     & 2018.02.27\\
TGBT17B\_502   & 01     & 2018.04.06\\
AGBT17A\_245   & 04     & 2018.04.06\\
AGBT17A\_245   & 05     & 2018.04.09\\
AGBT18B\_370   & 01     & 2018.12.09\\
AGBT18B\_370   & 02     & 2019.03.24\\
\hline
 \end{tabular}
\end{table}

\clearpage
\section{Appendix B}

The efficiencies are explicitly defined in their integral form
following the conventions of Kutner \& Ulich (1981).  The rearward
scattering and spillover efficiency is
\begin{equation}
\eta_{rss}= \frac{\int\!\!\int_{2\pi} P_{n}(\Omega)d\Omega}{\int\!\!\int_{4\pi} P_{n}(\Omega)d\Omega},
\end{equation}
where $P_{n}$ is the normalized antenna power pattern ($P_{n}(0)=1$)
and $2\pi$ is in the forward (on-sky) direction.  The efficiency
$\eta_{l} =\eta_{r}\eta_{rss}$, where the ohmic efficiency is
$\eta_{r}\simeq 1.0$.  The forward scattering and spillover efficiency
is
\begin{equation}
\eta_{fss}= \frac{\int\!\!\int_{\Omega_{D}} P_{n}(\Omega)d\Omega}{\int\!\!\int_{2\pi} P_{n}(\Omega)d\Omega},
\end{equation}
where $\Omega_{D}$ is the antenna diffraction pattern, including the
error pattern.

The main-beam efficiency is given by
\begin{equation}
\eta_{mb}= \frac{\int\!\!\int_{\Omega_{mb}} P_{n}(\Omega)d\Omega}{\int\!\!\int_{4\pi} P_{n}(\Omega)d\Omega},
\end{equation}
where $\Omega_{mb}$ is the main-beam lobe.  Assuming a Gaussian beam,
the integral $\int\!\!\int_{\Omega_{mb}} P_{n}(\Omega)d\Omega =
1.133\theta_{FWHM}^{2}$.  Although the main-beam lobe is well
approximated by a Gaussian, the antenna pattern at larger radii is
not.  For a circular aperture radio dish, the antenna pattern can be
represented by a Bessel function (e.g., Wilson, Rohlfs, \&
H\"{u}ttemeister 2010), and the relative strength of the main-beam
lobe to the side-lobes depend on the feed taper.  In general, the
standard integral limit for $\Omega_{mb}$ is at the diameter
corresponding to the beam width of the first null (BWFN) for computing
$\eta_{mb}$.  Assuming typical feed tapers used in radio astronomy and
adopting a Bessel function approximation for the beam pattern, $BWFN
\approx 2.7\theta_{FWHM}$.

The corrected main-beam efficiency is
\begin{equation}
\eta_{M}^{*}= \frac{\int\!\!\int_{\Omega_{mb}} P_{n}(\Omega)d\Omega}{\eta_{r}\int\!\!\int_{\Omega_{D}} P_{n}(\Omega)d\Omega},
\end{equation}
which is $\eta_{M}^{*} =\eta_{mb}/(\eta_{l}\,\eta_{fss})$.  For
sources larger the beam size, the measured beam efficiency depends on
the coupling of the source to the antenna beam pattern
(equation~10).  The coupling efficiency $\eta_{c}$ is 
\begin{equation}
\eta_{c} = \frac{\int\!\!\int_{\Omega_{\rm Source}} P_{n}(\Psi - \Omega)B_{n}(\Psi)d\Psi}{\int\!\!\int_{\Omega_{D}} P_{n}(\Omega)d\Omega},
\end{equation}
where $\Psi$ is the source angle on the sky and $B_{n}$ is the
normalized source brightness temperature ($B_{n}(0)=1$).  This
definition is consistent with Kutner \& Ulich (1981) where $T_{R}^{*}
= \eta_{c}\,T_{S}$.

\clearpage
\section{Figures}

\begin{figure}[tbh]
\begin{center}
\includegraphics[width=0.99\textwidth]{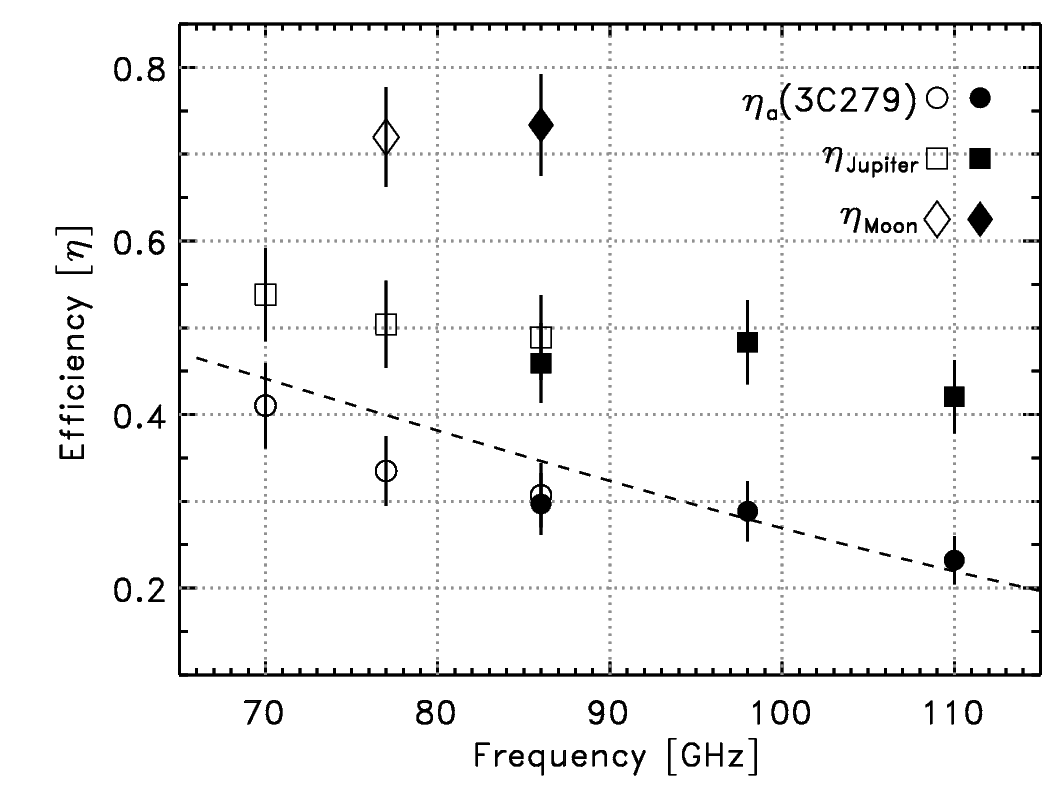}
\end{center}
\vspace*{-5mm}

\caption{The comparison of calibration measurements made with the
  W-band 4mm Receiver (open symbols) and Argus (solid symbols) within
  the same observing session (\S4.1).  The point-source aperture
  efficiencies are shown as circles, while the measured beam
  efficiencies of Jupiter and the Moon are shown as squares and
  diamonds, respectively.  The aperture efficiency based on the
  average good weather surface error of 235$\mu$m is shown as the
  dashed-line for comparison.}

\end{figure}

\clearpage
\begin{figure}[tbh]
\begin{center}
\includegraphics[width=0.99\textwidth]{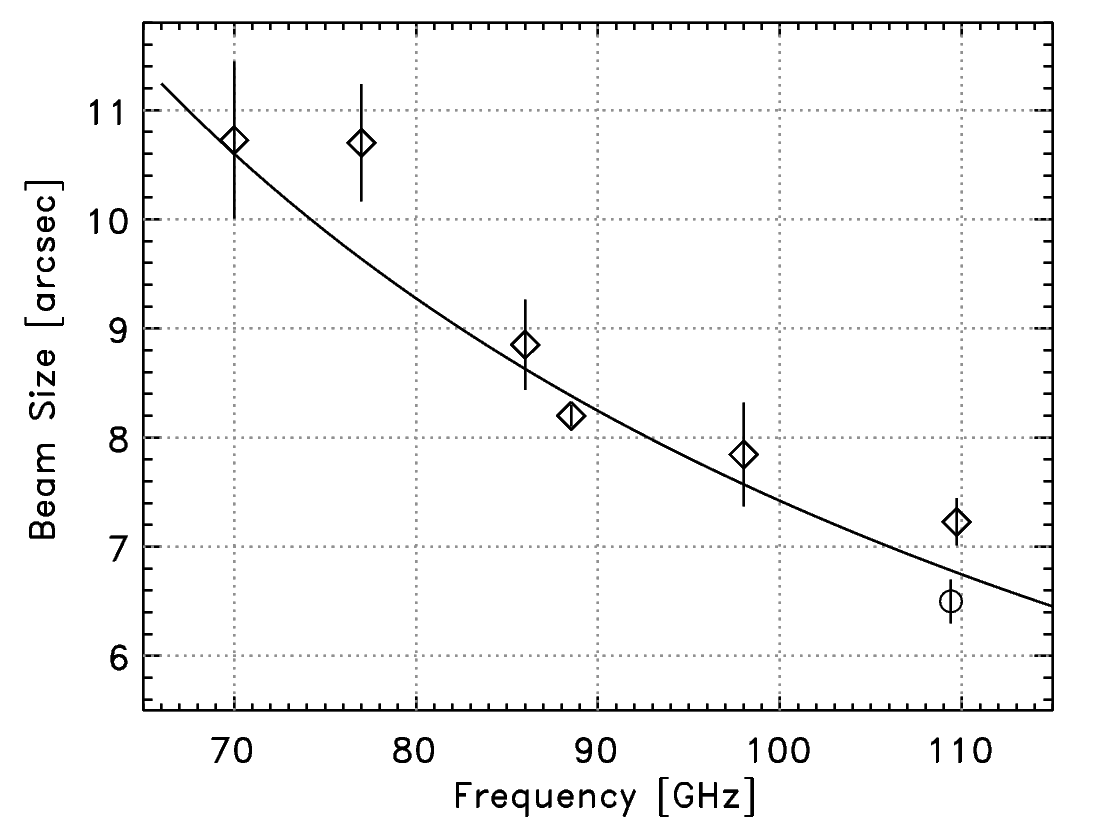}
\end{center}
\vspace*{-5mm}

\caption{The measured average $\theta_{\rm FWHM}$ beam sizes as a
  function of frequency shown as diamonds.  The solid line corresponds
  to $1.2 \lambda/D$.  On average there is some degradation of the
  measured beam size at highest frequencies ($\nu > 100$\,GHz), but
  the open circle shows the result at 109 GHz derived in excellent
  conditions (GBT Memo\#296) for comparison.}

\end{figure}

\clearpage
\begin{figure}[tbh]
\begin{center}
\includegraphics[width=0.99\textwidth]{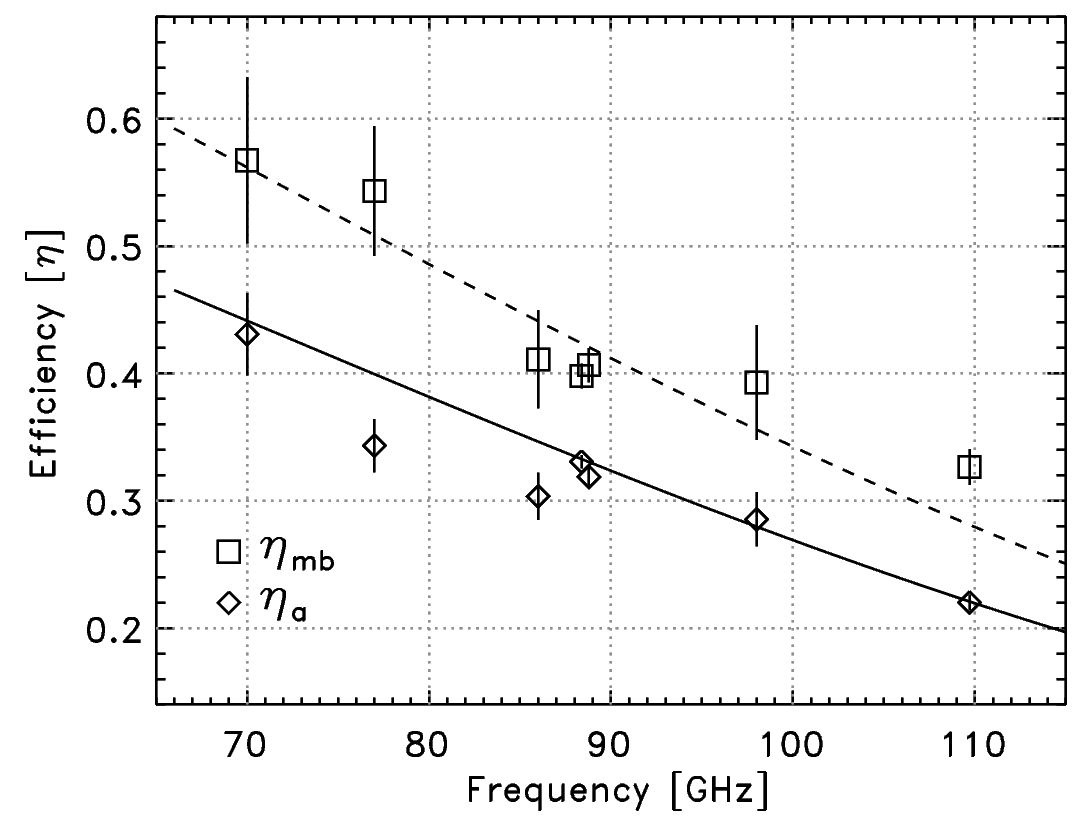}
\end{center}
\vspace*{-5mm}

\caption{The measured aperture and main-beam efficiencies as a
  function of frequency.  The diamonds represent the aperture
  efficiencies, while the squares are the measured main-beam
  efficiencies derived for point sources. The solid line shows the
  aperture efficiencies predicted by the Ruze equation with an
  effective surface error of 235$\mu$m.  The dashed line corresponds
  to an average ratio of $\eta_{mb}/\eta_{a} = 1.274$.}

\end{figure}

\clearpage

\begin{figure}[tbh]
\begin{center}
\includegraphics[width=0.99\textwidth]{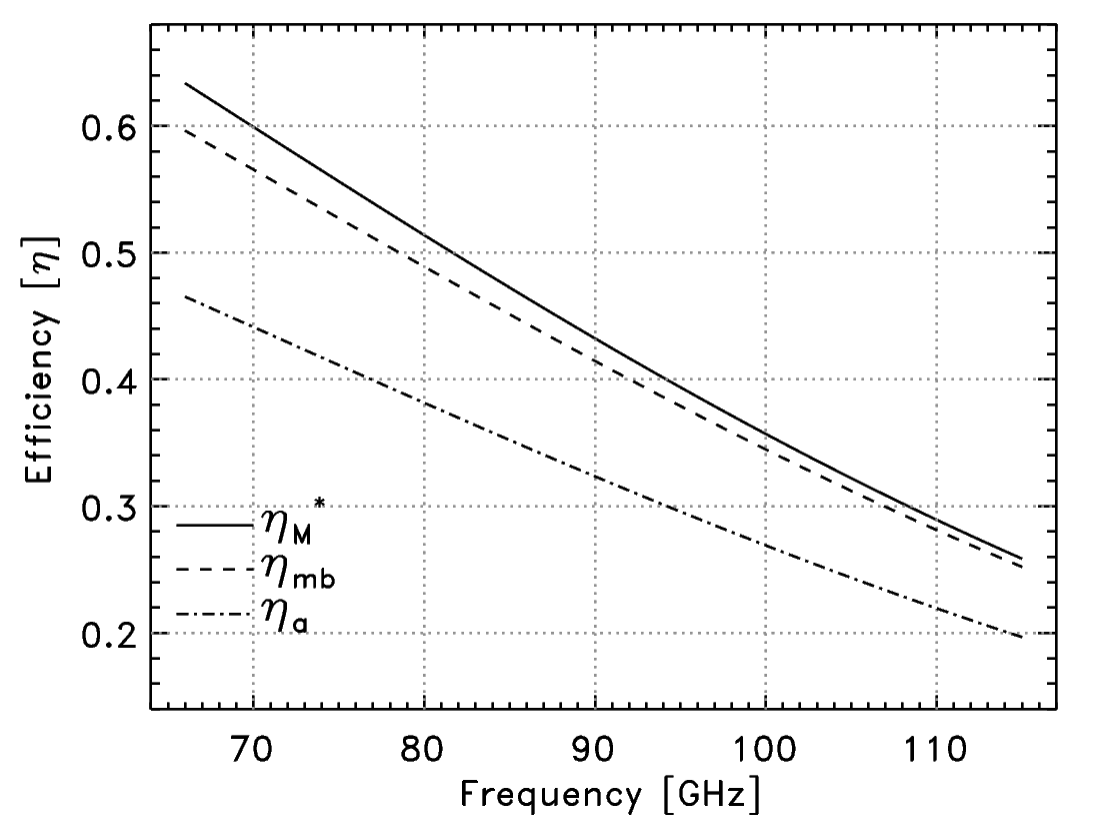}
\end{center}
\vspace*{-5mm}

\caption{The estimated efficiencies of the GBT as a function of
  frequency assuming an effective surface error of 235$\mu$m and a
  beam size parameter of $\kappa=1.2$.  The dashed-dotted line is the
  aperture efficiency.  The main-beam efficiency is shown by the
  dashed line, while the corrected main-beam efficiency is given by
  the solid line.}
\end{figure}

\clearpage
\begin{figure}[tbh]
\begin{center}
\includegraphics[width=0.99\textwidth]{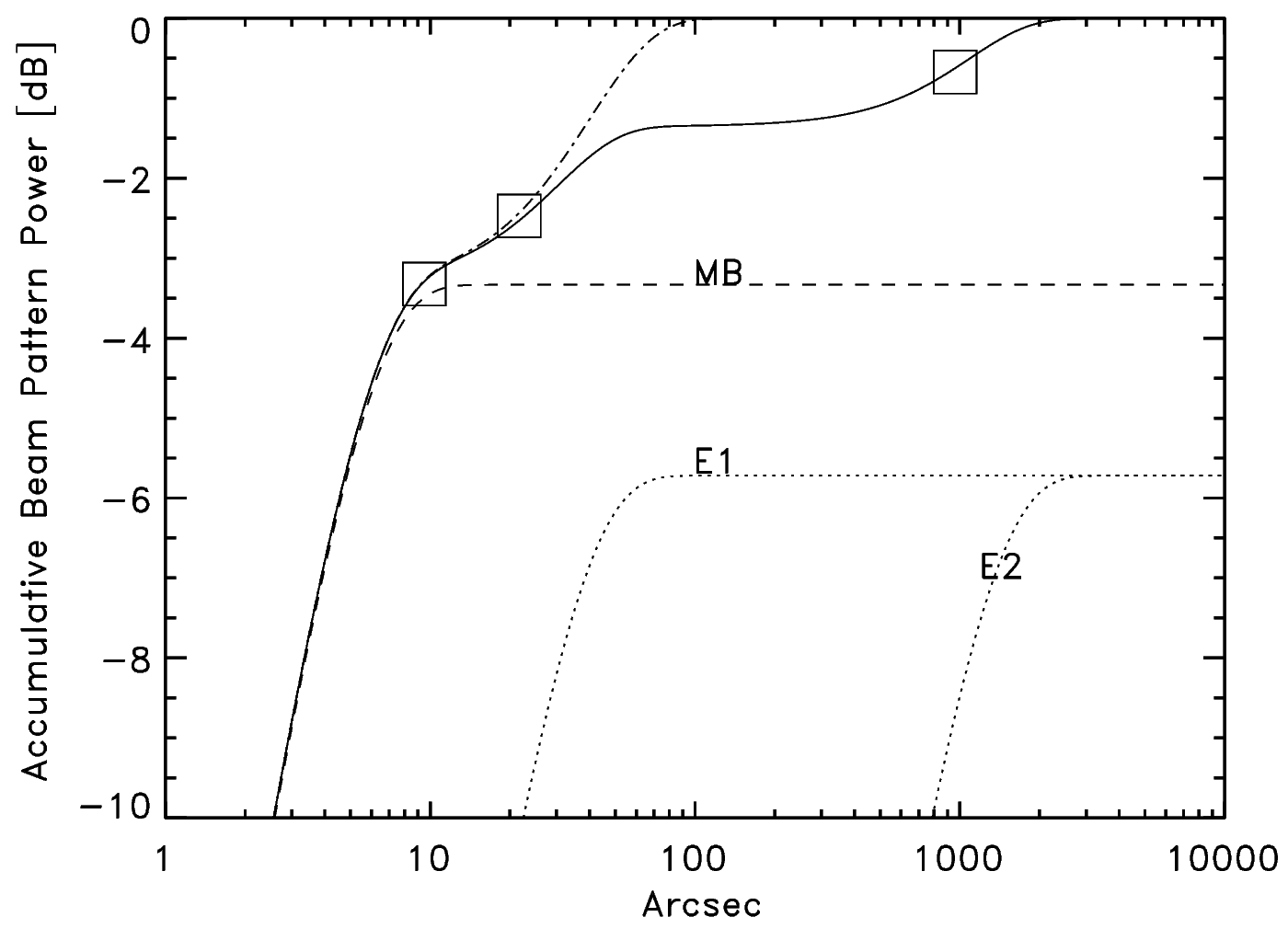}
\end{center}
\vspace*{-5mm}

\caption{The accumulative beam pattern power as a function of radius
  at 86 GHz.  The boxes show the measured beam efficiencies for
  small sources, Jupiter, and the Moon.  The dashed-dotted line is a
  fit assuming only one Gaussian component for the error pattern,
  while the solid line shows the fit using two Gaussian components for
  the error pattern.  The dashed and dotted lines show the
  accumulative power as a function radius for the individual
  components (main-beam [MB] and error-beam components [E1, E2], see
  Table~1).  The power is plotted in units of dB ($10\log_{10}[P]$).}

\end{figure}

\clearpage

\begin{figure}[htb]
\begin{center}
\includegraphics[width=0.99\textwidth]{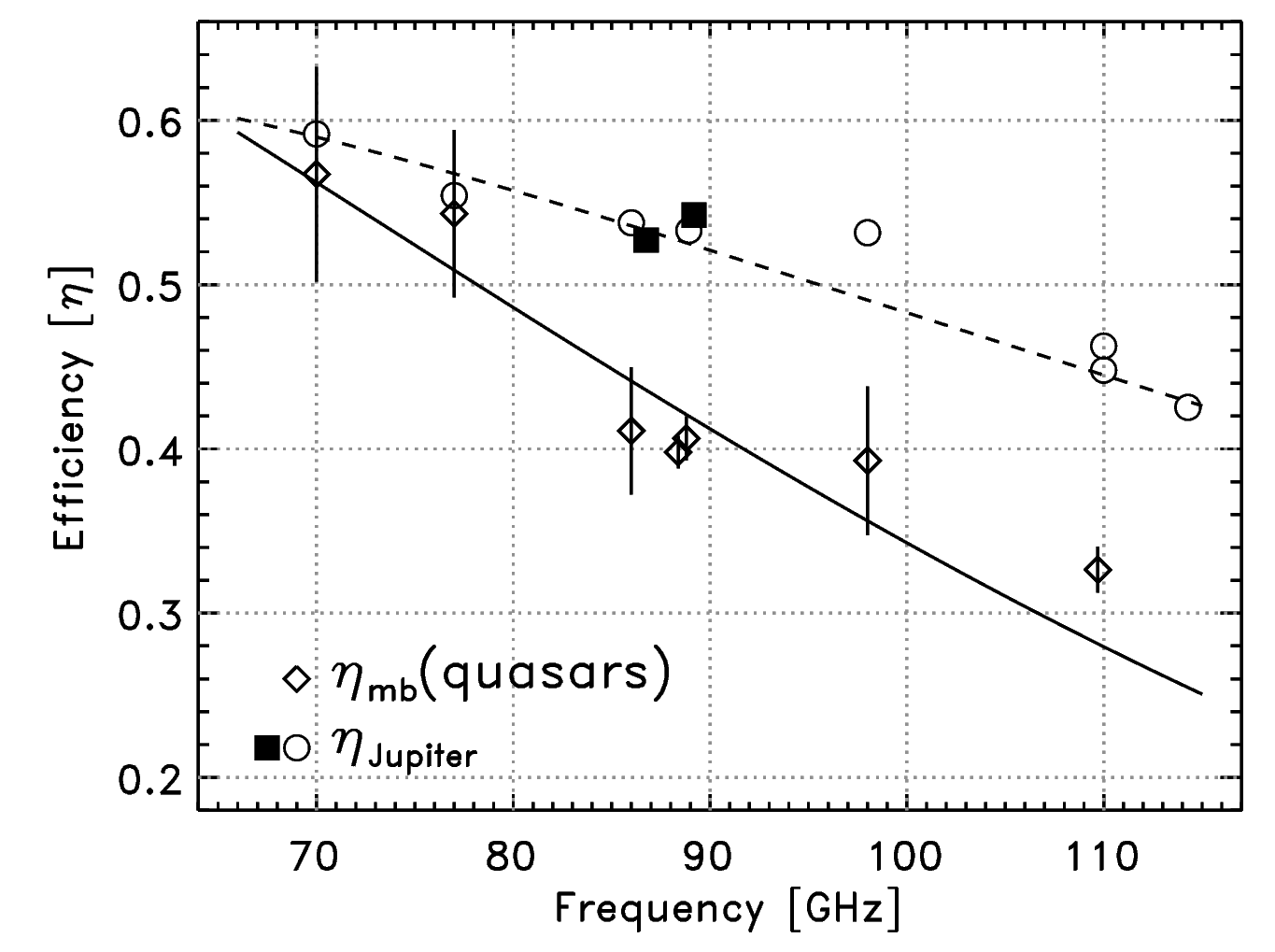}
\end{center}
\vspace*{-5mm}

\caption{The diamonds show the main-beam efficiencies measured using
  quasars, and the solid line corresponds to an average ratio of
  $\eta_{mb}/\eta_{a} = 1.274$.  The solid squares show measurements
  of Jupiter for two sessions when the telescope had good aperture
  efficiencies.  The circles represent $\eta_{\rm Jupiter}$ corrected
  for the relative aperture efficiencies between different sessions.
  The dashed line is an empirical relationship fitted to the circles
  and solid squares representing $\eta_{\rm Jupiter}$ as a function of
  frequency.}

\end{figure}

\end{document}